\documentclass[useAMS,usenatbib,usegraphicx]{mn2e}
\usepackage{epsfig}
\usepackage{appendix}
\usepackage{booktabs}
\usepackage{upgreek}
\usepackage{url}
\usepackage[usenames]{color}
\newcommand{\ion}[2]{{#1}\,{\sc #2}}

\usepackage{graphicx}
\usepackage{rotating}
\usepackage{color}
\usepackage{pifont}
\usepackage{pdflscape,lscape}
\usepackage{natbib}

\usepackage{txfonts,longtable}

\def\bz{\left< B_z\right>}
\newcommand{\CPD}{{CPD\,$-$62$^{\circ}$2124}}

\title[
Variability of the strongly magnetic CPD\,$-$62$^{\circ}$2124
]{
Rotationally modulated variability and pulsations of the He-rich star CPD\,$-$62$^{\circ}$2124 with an extraordinarily strong magnetic field
}

\author[
Hubrig et al.
]{
S.~Hubrig$^1$\thanks{E-mail: shubrig@aip.de},
Z.~Mikul\'a\v{s}ek$^2$,
A.~F.~Kholtygin$^3$,
I.~Ilyin$^1$,
M.~Sch\"oller$^4$,
S.~P.~J\"arvinen$^1$,
\and R.-D.~Scholz$^1$,
M.~Zejda$^2$\\
$^1$Leibniz-Institut f\"ur Astrophysik Potsdam (AIP), An der Sternwarte~16, 14482~Potsdam, Germany\\
$^2$Department of Theoretical Physics and Astrophysics, Masaryk University, Kotl\'a\v{r}sk\'a 2, 611~37~Brno, Czech Republic\\
$^3$Saint-Petersburg State University, Universitetskij pr.~28, Saint-Petersburg 198504, Russia\\
$^4$European Southern Observatory, Karl-Schwarzschild-Str.~2, 85748~Garching, Germany
}

\begin{document}

\def\teff{${T}_{\rm eff}$}
\def\kms{{km\,s}$^{-1}$}
\def\logg{$\log g$}
\def\micro{$\xi_{\rm t}$}
\def\macro{$\zeta_{\rm RT}$}
\def\rad{$v_{\rm r}$}
\def\vsini{$v\sin i$}
\def\ebv{$E(B-V)$}

\date{Accepted ... Received ...; in original form ...}

\pagerange{\pageref{firstpage}--\pageref{lastpage}} \pubyear{2002}

\maketitle

\label{firstpage}

\begin{abstract}
A longitudinal magnetic field with a strength of 5.2\,kG was recently detected
in CPD\,$-$62$^{\circ}$2124, which has a fractional main-sequence lifetime of about 60\%.
Strongly magnetic early-B type chemically peculiar stars in an advanced evolutionary state
are of special interest to understand the evolution of the angular momentum
and spin-down timescales in the presence of a global magnetic field.
We exploited 17 FORS\,2 low-resolution spectropolarimetric observations and 844 ASAS3 photometric measurements
for the determination of the rotation period, pulsations,
and the magnetic field geometry of the star.
We calculated periodograms and applied phenomenological models
of photometric, spectral and spectropolarimetric variability.
We found that all quantities studied, specifically equivalent widths,
the mean longitudinal magnetic field $\bz{}$, and the flux in the $V$ filter,
vary with the same period $P = 2.628$\,d, which was identified as the rotation period.
The observed variations can be fully explained by a rigidly rotating main-sequence star
with an uneven distribution of chemical elements, photometric spots, and a stable,
nearly dipolar magnetic field with a polar field strength of about 21\,kG, frozen into the body of the star.
The magnetic field of CPD\,$-$62$^{\circ}$2124 is
tilted to the rotation axis by $\beta=28^{\circ}\pm7^{\circ}$,
while the inclination of the rotation axis towards the line of sight is only $i=20^{\circ}\pm5^{\circ}$.
In the acquired FORS\,2 spectra, we detect short-term line profile variations indicating the presence of $\beta$~Cephei 
type pulsations.
As of today, no other pulsating star of this type is known to possess such a strong magnetic field.
\end{abstract}


\section{Introduction}
\label{sect:intro}

The detection of a very strong longitudinal magnetic field of the order of 5.2\,kG
in the chemically peculiar early-B type star \CPD{}
was previously reported by \citet{Castro2017} who used one spectropolarimetric observation with the High Accuracy Radial
velocity Planet Searcher (HARPS\-pol; \citealt{Snik2011}) attached to the ESO 3.6m telescope on La~Silla and one
observation with the FOcal Reducer low dispersion
Spectrograph (FORS\,2; \citealt{Appenzeller1998}) mounted on the 8\,m Antu telescope of the VLT.
Assuming a dipolar
magnetic field geometry to estimate the magnetic field strength of about 18.3\,kG,
\citet{Castro2017}
suggested that no other early-B type star is known to host a magnetic field with a comparable strength.
The authors also reported that this star exhibits strong, broad H$\alpha$ emission that is characteristic of a centrifugal magnetosphere.
Moreover, a study of the stellar parameters, using a comparison with stellar evolutionary tracks,
indicated that \CPD{} is largely evolved from the zero-age main sequence with a fractional main-sequence lifetime of about 60\%.

Our recent monitoring of this star employing multi-epoch FORS\,2 spectropolarimetric observations
distributed over about two and a half months revealed that the magnetic field is even stronger than previously anticipated,
with a mean longitudinal magnetic field strength $\bz{}$, varying from 4.0 to 6.8\,kG.

In the following sections, we present the results of our search for rotationally and pulsationally modulated variability
of 844 photometric ASAS3 data points, the magnetic field measurements,
and the equivalent widths (EWs) of nine lines belonging to six chemical elements.

\section{Observations}
\label{sect:obs}

\subsection{ASAS3 photometry}

We have used archival data from the ASAS3 survey\footnote{http://www.astrouw.edu.pl/asas/},
covering the time interval from HJD 2\,451\,888.793 (2000 December 10) to 2\,455\,048.516 (2009 August 5).
After removing some apparent outliers, we used 844 data points in the $V$ band.
These measurements were made by the ASAS3 system \citep{pac00}, which consists of two wide-field 200/2.8 instruments,
one narrow-field 750/3.3 telescope and one super-wide 50/4 scope, each equipped with an Apogee 2K\,$\times$\,2K CCD camera,
located at Las Campanas Observatory, Chile (since 1997) and on Haleakala, Maui (since 2006).

\subsection{Spectropolarimetry}

Sixteen FORS\,2 spectropolarimetric observations of CPD\,$-$62$^{\circ}$2124 were obtained
from 2016 March~12 to May~24.
The FORS\,2 multi-mode instrument is equipped with polarisation analysing optics
comprising super-achromatic half-wave and quarter-wave phase retarder plates,
and a Wollaston prism with a beam divergence of 22$\arcsec$ in standard
resolution mode.
We used the GRISM 600B and the narrowest available slit width
of 0$\farcs$4 to obtain a spectral resolving power of $R\sim2000$.
The observed spectral range from 3250 to 6215\,\AA{} includes all Balmer lines,
apart from H$\alpha$, and numerous helium lines.
For the observations, we used a non-standard readout mode with low
gain (200kHz,1$\times$1,low), which provides a broader dynamic range, hence
allowing us to reach a higher signal-to-noise ratio (SNR) in the individual spectra.
The exposure time for each subexposure
accounted for 7\,min. A detailed description of the assessment of longitudinal magnetic field
measurements using FORS\,1/2 spectropolarimetric observations was presented in detail
in our previous work (e.g., \citealt{Hubrig2014,Hubrig2017}, and references therein).

\begin{table}
\caption{
Logbook of the FORS\,2 polarimetric observations of \CPD{}, including
the modified Julian date of mid-exposure followed by the
achieved signal-to-noise ratio in the Stokes~$I$ spectra around 5000\,\AA{},
and the measurements of the mean longitudinal magnetic field using the
Monte Carlo bootstrapping test, only for the hydrogen lines and for all lines.
In the last column, we present the results of our measurements using the null spectra for the set of all lines.
All quoted errors are 1$\sigma$ uncertainties.
}
\label{tab:Bz}
\centering
\begin{tabular}{lcr@{$\pm$}rr@{$\pm$}rr@{$\pm$}r}
\hline
\hline
\multicolumn{1}{c}{MJD} &
\multicolumn{1}{c}{SNR} &
\multicolumn{2}{c}{$\left< B_{\rm z}\right>_{\rm hyd}$} &
\multicolumn{2}{c}{$\left< B_{\rm z}\right>_{\rm all}$} &
\multicolumn{2}{c}{$\left< B_{\rm z}\right>_{\rm N}$} \\
&
&
 \multicolumn{2}{c}{[G]} &
 \multicolumn{2}{c}{[G]} &
 \multicolumn{2}{c}{[G]} \\
\hline
     57099.2449 &1800 & 4640 &  130  & 4530&   100  &  10  &  80 \\
     57460.1011 &1100 & 5600 &  300  & 5510&  180  &$-$10  & 150 \\
     57461.0599 &1300 & 6500 &  300  & 6270&  160  &$-$40 & 100 \\
     57462.0660 &1400 & 3970 &  140  & 3740&   100  &   20 &  90 \\
     57463.0650 &1300 & 5600 &  300  & 5700&  150  &    0 & 120 \\
     57465.2482 &1400 & 5100 &  200  & 4670&  180  &  140 & 130 \\
     57474.0863 &1000 & 6800 &  500  & 5600&  200  &$-$40 & 170 \\
     57478.2042 &1700 & 4470 &  150  & 4120&  120  &$-$20 &  90 \\
     57478.2560 &1700 & 4270 &  180  & 4140&  120  &   10 &  90 \\
     57478.3208 &1500 & 4800 &  190  & 4590&  150  &  150 & 110 \\
     57494.0163 &1400 & 3940 &  180  & 4140&  130  &  $-$10& 100 \\
     57495.0476 &1200 & 5150 &  250  & 5520&  140  &     0& 110 \\
     57500.1136 &1700 & 5040 &  160  & 5290&   100  & $-$60&  70 \\
     57505.2651 &1300 & 5050 &  250  & 5480&  140  &    60& 110 \\
     57530.1287 &1300 & 5080 &  200  & 4050&  250  &   10 & 170 \\
     57531.2140 &1100 & 5900 &  400  & 5500&  170  &$-$250& 130 \\
     57532.0571 &1300 & 6650 &  250  & 6180&  120  & $-$30&  90 \\
\hline
\end{tabular}
\end{table}

The longitudinal magnetic field was measured in two ways: using the entire spectrum
including all available lines, or using exclusively the hydrogen lines.
Furthermore, we have carried out Monte Carlo bootstrapping tests (e.g.\ \citealt{Hubrig2014,Hubrig2015a}).
These are most often applied with the purpose of deriving robust estimates of standard errors.
The measurement uncertainties obtained with and without the Monte Carlo bootstrapping tests were found to be
in close agreement, indicating the absence of reduction flaws.
The results of our magnetic field measurements, those for the entire spectrum
and those only for the hydrogen lines are presented in
Table~\ref{tab:Bz}, where we also include in the first row the information about the previous
magnetic field measurement with FORS\,2 presented by \citet{Castro2017},
who also presented in their work a typical example of
the analysis of the FORS\,2 data.

\begin{table*}
\begin{center}
\caption{
Equivalent widths of nine spectral lines belonging to different elements, measured on 17 FORS\,2 spectrograms,
MJD is the modified Julian date of mid-exposure.
}
\label{tab:EW}
\centering
\begin{tabular}{rrrrrrrrrr}
\hline
\hline
\multicolumn{1}{c}{MJD} &
\multicolumn{9}{c}{Equivalent widths} \\
 &
\multicolumn{1}{c}{H$\beta$} &
\multicolumn{1}{c}{H$\gamma$} &
\multicolumn{1}{c}{\ion{He}{i}\,4388} &
\multicolumn{1}{c}{\ion{He}{i}\,4471} &
\multicolumn{1}{c}{\ion{He}{i}\,4922} &
\multicolumn{1}{c}{\ion{C}{ii}\,4267} &
\multicolumn{1}{c}{\ion{N}{ii}\,4631} &
\multicolumn{1}{c}{\ion{O}{ii}\,4662} &
\multicolumn{1}{c}{\ion{Si}{iii}\,4553} \\
 &
\multicolumn{1}{c}{[\AA{}]} &
\multicolumn{1}{c}{[\AA{}]} &
\multicolumn{1}{c}{[\AA{}]} &
\multicolumn{1}{c}{[\AA{}]} &
\multicolumn{1}{c}{[\AA{}]} &
\multicolumn{1}{c}{[\AA{}]} &
\multicolumn{1}{c}{[\AA{}]} &
\multicolumn{1}{c}{[\AA{}]} &
\multicolumn{1}{c}{[\AA{}]} \\
\hline
57099.2449 &   3.0456 &    3.3339 &  1.6523 & 2.4379 &   1.4058 &  0.1772 &  0.0766 &  0.0640  &  0.1805 \\
57460.1011 &   2.5237 &    2.8773 &  1.9360 & 2.7332 &   1.5604 &  0.1550 &  0.0489 &  0.0487  &  0.1511 \\
57461.0599 &   2.1685 &    2.5917 &  2.1532 & 2.8583 &   1.6600 &  0.1103 &  0.0341 &  0.0399  &  0.1219 \\
57462.0660 &   3.0769 &    3.4358 &  1.6156 & 2.4379 &   1.3734 &  0.1744 &  0.0636 &  0.0569  &  0.1916 \\
57463.0650 &   2.0480 &    2.5788 &  2.1665 & 2.8861 &   1.6642 &  0.1172 &  0.0329 &  0.0306  &  0.1255 \\
57465.2482 &   2.6128 &    3.1209 &  1.8857 & 2.6540 &   1.5343 &  0.1571 &  0.0583 &  0.0596  &  0.1800 \\
57474.0863 &   2.0102 &    2.5956 &  2.1980 & 2.8318 &   1.6582 &  0.0981 &  0.0478 &  0.0316  &  0.1320 \\
57478.2042 &   2.9142 &    3.2343 &  1.7594 & 2.5529 &   1.4687 &  0.1593 &  0.0420 &  0.0755  &  0.1919 \\
57478.2560 &   2.8194 &    3.1810 &  1.7842 & 2.5660 &   1.4848 &  0.1541 &  0.0409 &  0.0709  &  0.1860 \\
57478.3208 &   2.7317 &    3.1318 &  1.8634 & 2.5681 &   1.5084 &  0.1425 &  0.0489 &  0.0692  &  0.1896 \\
57494.0163 &   2.8598 &    3.1215 &  1.7442 & 2.5013 &   1.4012 &  0.1634 &  0.0572 &  0.0662  &  0.1853 \\
57495.0476 &   2.0469 &    2.5635 &  2.2016 & 2.8957 &   1.6833 &  0.1015 &  0.0521 &  0.0468  &  0.1144 \\
57500.1136 &   2.0258 &    2.5119 &  2.2482 & 2.9045 &   1.7024 &  0.1073 &  0.0288 &  0.0516  &  0.0823 \\
57505.2651 &   1.9919 &    2.7024 &  2.2178 & 2.8703 &   1.6980 &  0.1083 &  0.0356 &  0.0293  &  0.1067 \\
57530.1287 &   2.9581 &    3.2974 &  1.6719 & 2.5204 &   1.3991 &  0.1717 &  0.0699 &  0.0734  &  0.1741 \\
57531.2140 &   2.2194 &    2.7839 &  2.0940 & 2.8458 &   1.6821 &  0.1247 &  0.0422 &  0.0651  &  0.1303 \\
57532.0571 &   2.1657 &    2.7054 &  2.1139 & 2.8611 &   1.6390  & 0.1123  & 0.0402  & 0.0549  &  0.1230 \\
\hline
\end{tabular}
\end{center}
\end{table*}
                                                                                                                                           
Furthermore, to search for the presence of spectral variability, the FORS\,2 Stokes~$I$ spectra were used to measure the EWs
of nine spectral lines:
H$\beta$, H$\gamma$, \ion{He}{i}~4388, \ion{He}{i}~4471, \ion{He}{i}~4922, \ion{C}{ii}~4267, \ion{N}{ii}~4631,
\ion{O}{ii}~4662, and \ion{Si}{iii}~4553. The measurements were carried out using the gravity method
and are presented in Table~\ref{tab:EW}.

\section{Search for periodic variability}
\label{sect:pervar}                                                                                                                        
\CPD{} is classified as a hot magnetic star \citep{Castro2017},
characterized by an enhanced surface abundance of helium,
which is usually concentrated in large spots surviving for many decades \citep{mik16}.
The surface distribution of some other chemical elements, like silicon or carbon, also displays a spotted structure, which is,
as a rule, disjunct from the helium distribution
\citep[see e.g.\ the well-studied He-rich star HD\,37776;][]{chochlo}.
The enhanced redistribution of radiative flux from the ultraviolet to the visible region of hot CP stars
causes the presence of large bright photometric spots, usually with overabundant silicon,
which is the most efficient, having plenty of bound-free transition continua in the ultraviolet \citep{krt07,krt15}.

The global magnetic dipole-like field, tilted to the rotation axis by the angle $\beta$,
frozen in the plasma in the outer layers of the star, is also a long-lived phenomenon firmly bound with the rotating stellar surface.
As the star rotates, we should detect variations in the intensity of the longitudinal magnetic field $\bz{}$,
the spectral line profiles of various elements, and the photometric changes, all periodical with the rotation period
of the star.

There are some constraints on the rotation period of \CPD{}, following from the adapted model
and based on the observed projected rotation velocity and assumed dimension of the star corresponding to its effective temperature,
surface gravity, $v \sin i=35\pm5$\,km\,s$^{-1}$, $T_{\mathrm{eff}}=23\,650$\,K, and $\log\,g=3.95$\, according to \citet{Castro2017}.
The two competing evolutionary models (differing mostly in the value of the adopted overshooting parameter)
of \citet{Ekstroem2012} and \citet{Brott2011} yielded different estimates for the radius of \CPD{}:
$R=5.8\pm0.9$\,R$_{\odot}$ and $R=5.1\pm0.8$\,R$_{\odot}$, respectively.
For the following considerations, we adopt a compromise: $R=5.4\pm1.0\,R_{\odot}$.

The rotation period $P$ (in days) follows the relation: $P=50.6\, R\,\sin i/v \sin i=8(2)\,\sin(i)$\,d,
where $i$ is the inclination of the rotation axis and $R$ is the radius in R$_{\odot}$.
For the maximum inclination $i=90^{\circ}$ we obtain the maximum period $P_{\mathrm{max}}=8\pm2$\,d \citep[see also][]{Castro2017}.
Obviously, the rotation period can not be arbitrarily shorter than $P_{\mathrm{max}}$
because we observe apparent, mutually correlated/anticorrelated variations in spectral line intensities.
Such variations can be naturally explained by a model of a rotating star with spots with enhanced abundances
of some chemical elements, provided that the star is not nearly pole-on.

\begin{figure*}
  \centering
  \includegraphics[width=0.80\textwidth]{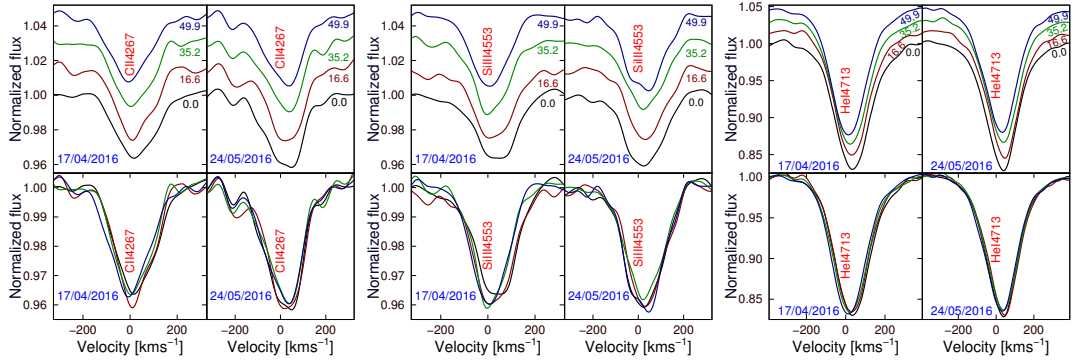}
\caption{
The behaviour of \ion{C}{ii}~4267, \ion{Si}{iii}~4553, and \ion{He}{i}~4713 (from left to right) in the FORS\,2  spectra
in each individual subexposure belonging to observations on two different epochs. For each epoch, in the upper row,
we present the line profiles shifted in vertical direction for best visibility.
The time difference (in minutes) between the subexposure and the start of
observations is given close to each profile. The lower row shows all profiles
overplotted.
}
\label{fig:puls_c}
\end{figure*}

We would also like to note that since CPD\,$-$62$^{\circ}$2124 already finished a fractional main-sequence lifetime of
about 60\% and is
already passing the $\beta$~Cep instability strip, the presence of shorter periods of the orders of a couple of hours attributed
to stellar pulsations can also be expected in our data. Indeed, in the acquired FORS\,2 spectra, along with different radial
velocity shifts of lines belonging to different  elements, we also detect
distinct changes in line profiles taking place on time-scales corresponding to the duration of the
sub-exposure sequences for the individual observations. In Fig.~\ref{fig:puls_c}
we present the behaviour of the line profiles in the individual spectral lines belonging to carbon, silicon, and helium.
The time difference between the individual subexposures was between 15 and 19\,min.

\subsection{Photometric variations}
\label{sect:fotovar}

Our period analysis of the moderately faint \CPD{} ($V= 10.432\,4(7)$\,mag)
is based on 844 ASAS3 $V$ measurements obtained during nine seasons between 2001 and 2009.
After removing some long-term trends, we have found that the relatively small scatter of data of 0.019\,mag
and the relative homogeneity allow us to search for periodic variations with amplitudes up to 0.008\,mag.
For the frequency analysis we have used, besides standard tools, also our own periodograms,
displaying amplitudes of sinus-like variations in magnitudes and other diagnostic tools,
published in \citet{pau13} and \citet{mik15}.

\begin{figure*}
\centering
\includegraphics[width=0.85\textwidth]{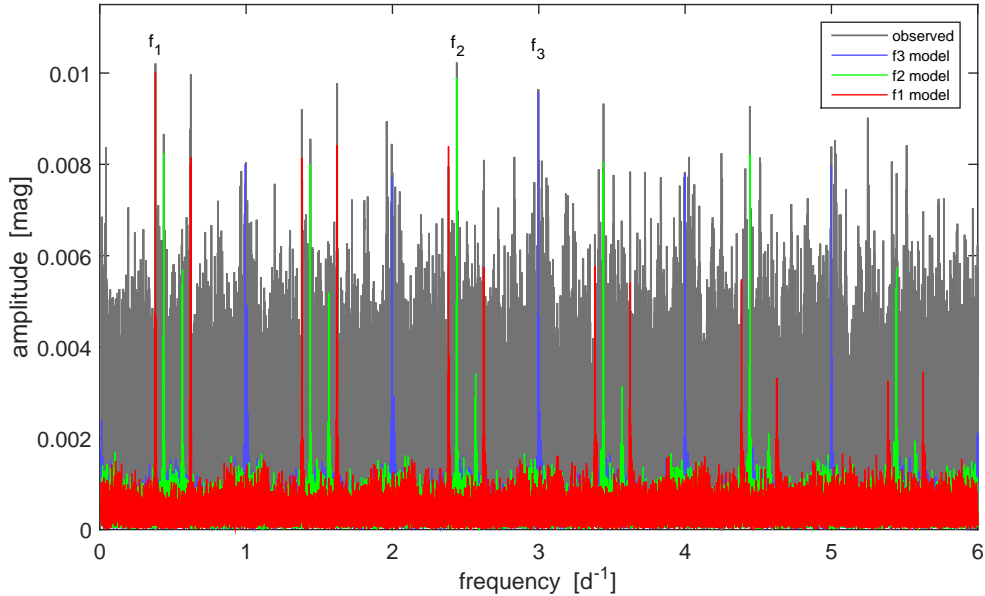}
\begin{center}
\caption{
The observed frequency spectrum of photometric data for periods larger than four hours (gray line in the online version 
of the article)
displays three independent peaks with amplitudes around 0.01\,mag.
The predicted frequency spectra of the individual components following Eq.~\ref{eq:fotmodel}
for the three frequencies $f_1$, $f_2$, and $f_3$
(red, green, and blue in the online version of the article),
also illustrate the position of adjacent aliases with respect to the principal modes.
}
\label{fig:Fotopermod}
\end{center}
\end{figure*}

Inspecting the frequency spectrum of \CPD{} in the ASAS3 data with frequencies below 6\,d$^{-1}$,
we revealed three weak, independent (non-aliased), but quite significant peaks at the frequencies
$f_1=0.3805\,\mathrm{d}^{-1}$,
$f_2=2.4397\,\mathrm{d}^{-1}$, and
$f_3=2.9974\,\mathrm{d}^{-1}$,
with amplitudes of 0.010\,mag (see Fig.~\ref{fig:Fotopermod}).
The significance of the complete frequency triple is 96.5\%.
The first period can be attributed to rotation, while the other two are probably pulsational periods.

The light curves plotted for the three respective periods are simple sinusoids and can be modeled by a simple relation:

\begin{equation}
\label{eq:fotmodel}
m(t)=m_0+\sum_{i=1}^3\frac{A_i}{2}\,\cos(2\,\pi\,\vartheta_i);\quad \vartheta_i(t)=\frac{t-M_{0i}}{P_i},
\end{equation}

\noindent
where $m(t)$ is the $V$ magnitude predicted for time $t$,
and for each of the three components $i$,
$A_i$, $P_i$, $M_{0i}$, and $\vartheta_i(t)$
are the amplitude, period, HJD of the basic maximum, and the value of the phase function for time $t$.
The frequency spectrum showing each component is presented in Fig.~\ref{fig:Fotopermod}.

The model frequency spectrum shows that the majority of the other observed significant peaks
can be explained as aliases of the three dominant frequency modes.
Considering that the maximum amplitude peaks caused by the random scatter have an amplitude of 0.083\,mag,
we can conclude that the model of photometric variability given by Eq.~(\ref{eq:fotmodel}) is quite
adequate for the available photometric observations.

\subsection{Variations of the equivalent widths of selected spectral lines and of the mean longitudinal magnetic field}

Information on the periodic components of the variability of \CPD{} is also hidden in the spectropolarimetric data
extracted from the 17 low-dispersion FORS\,2 spectrograms:
the EWs of the best unblended lines belonging to the elements H, He, C, N, O, and Si (see Table~\ref{tab:EW})
and the mean longitudinal magnetic field $\langle B_z\rangle$ (see Table~\ref{tab:Bz}).

\begin{figure}
\centering
\includegraphics[width=0.45\textwidth]{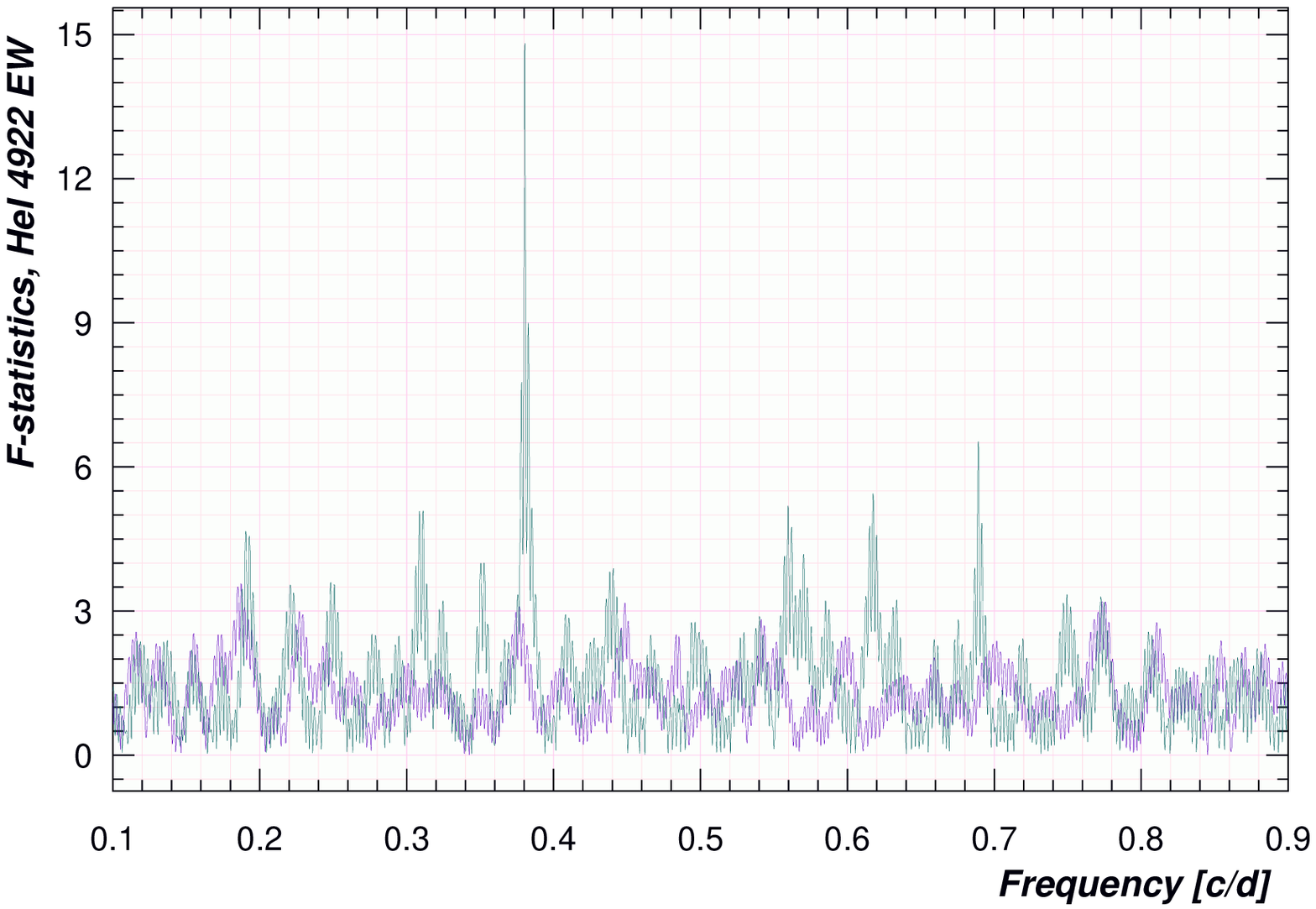}
\includegraphics[width=0.45\textwidth]{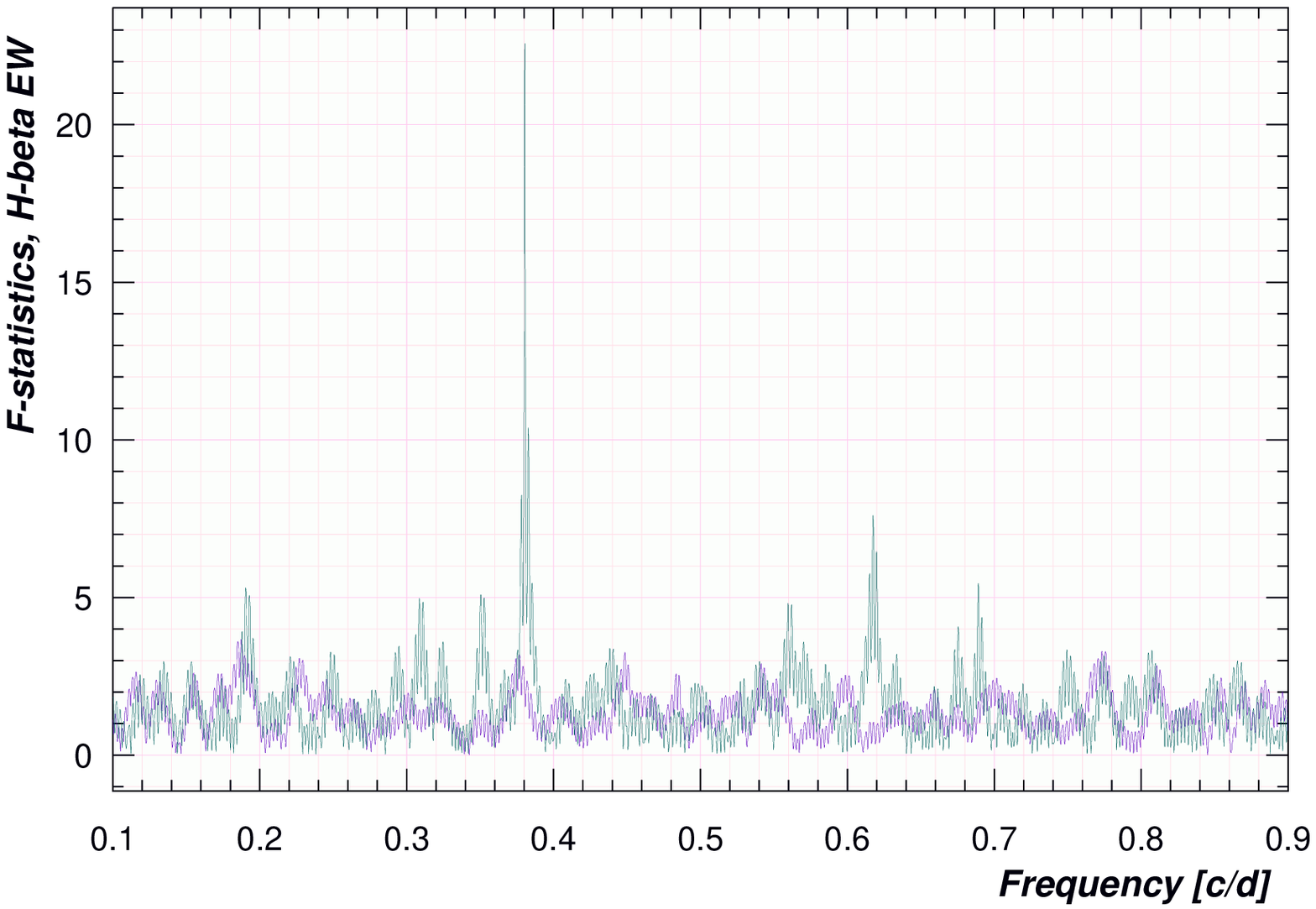}
\caption{
{\it Upper panel}:
F-statistics frequency periodogram (in d$^{-1}$) for the EWs of the \ion{He}{i} 4922 line.
{\it Lower panel}:
The same periodogram for the EWs of H$\beta$.
The window function is indicated by the blue colour in the online version of the article.
Both periodograms display only one dominant peak at the frequency 0.380\,d$^{-1}$.
}
\label{fig:hehper}
\end{figure}

The measured EW values of all nine selected spectral lines
and of the mean longitudinal magnetic field $\bz{}$
exhibit an apparent rotational modulation that one can expect in the case of a He-rich star with chemical spots
and a dipole-like magnetic field tilted to the rotation axis by a non-zero angle $\beta$.
It can be well illustrated by the F-statistics frequency spectra \citep{Seber1977}
obtained for the measurements of EWs of \ion{He}{i} 4922 and H$\beta$ (see Fig.~\ref{fig:hehper}).
The periodograms cover the whole frequency interval restricted for possible rotation frequencies.
All periodograms calculated for the EWs and $\bz{}$ exhibit only the period that coincides
with $P_1=2\fd628\,07(24)$, extracted from the photometry of the star.

A detailed inspection of the relationships between the EWs of the individual lines
and the $\bz{}$ measurements derived from the 17 FORS\,2 spectrograms shows that all quantities vary
and all variations are strongly mutually correlated or anticorrelated.
In particular the intensity of the \ion{He}{i} lines and the strength of the mean longitudinal magnetic field $\bz{}$
vary opposite to the intensity of the \ion{H}{i}, \ion{C}{ii}, \ion{N}{ii}, \ion{O}{ii}, and \ion{Si}{iii} lines.
Such a behavior can be explained if the center of a spot with enhanced helium abundance
is located in the vicinity of the north magnetic pole,
while the other elements are more abundant outside of the He spot.

\begin{figure}
\centering
\includegraphics[width=0.55\textwidth]{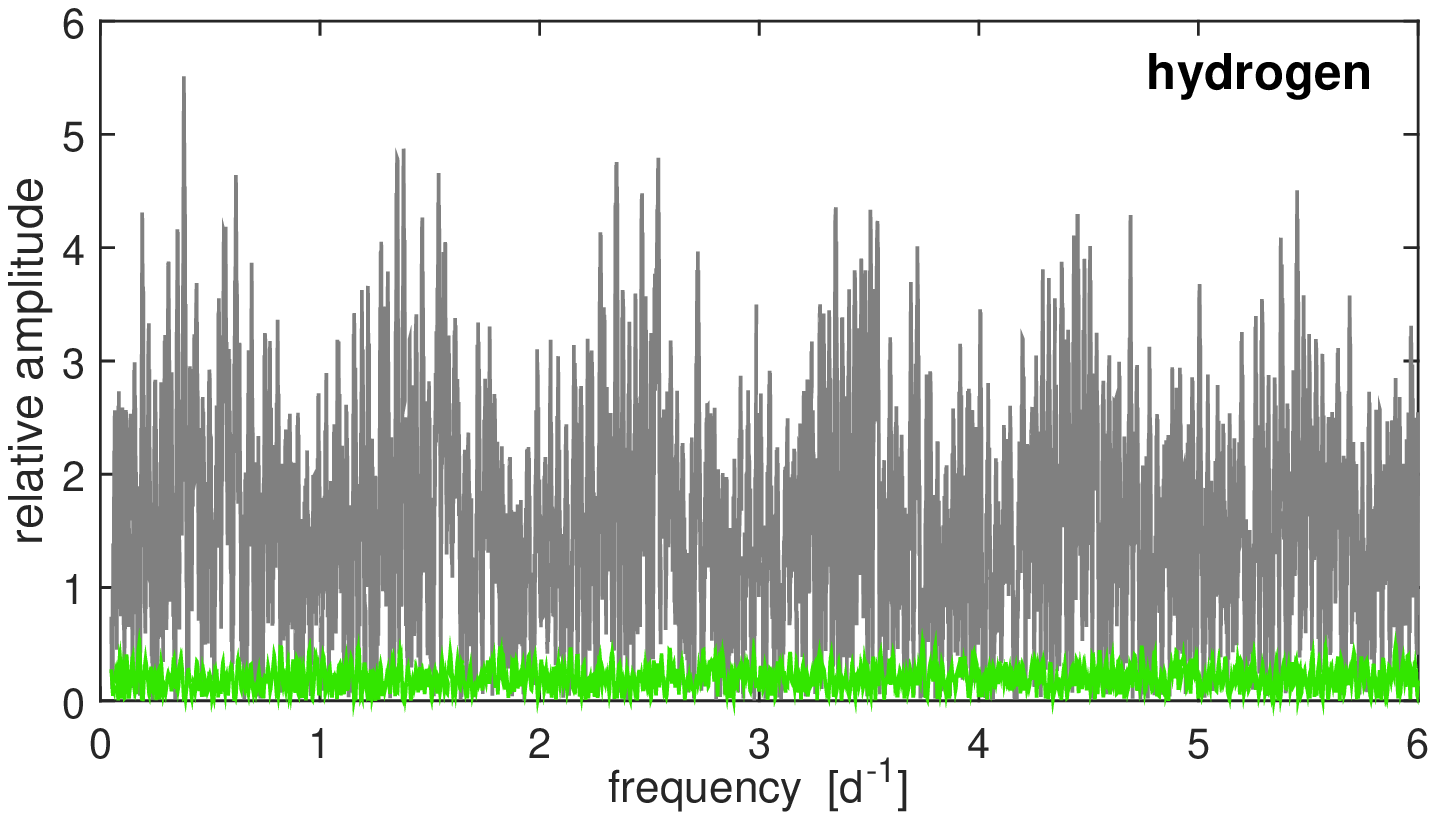}
\includegraphics[width=0.55\textwidth]{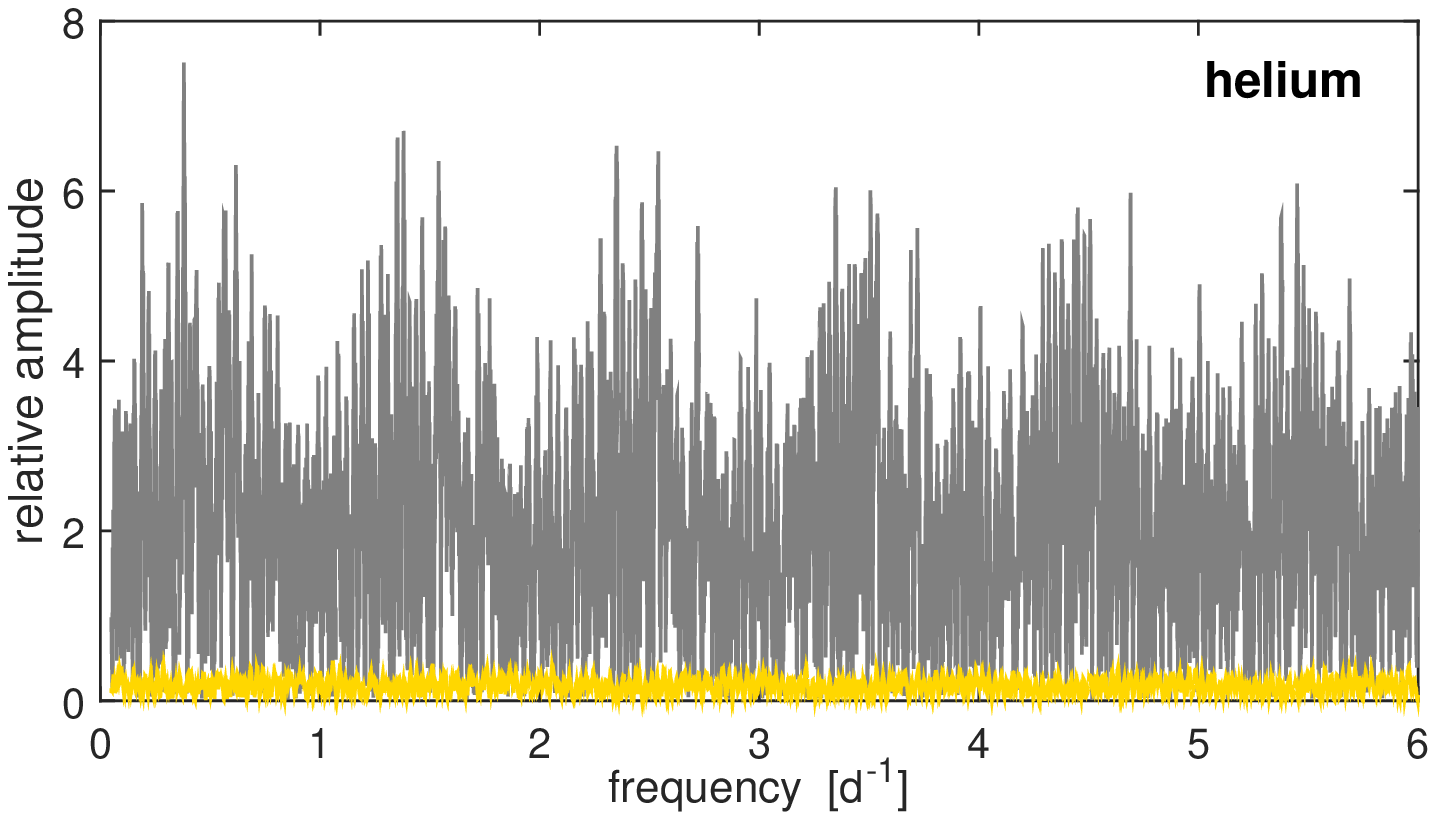}
\includegraphics[width=0.55\textwidth]{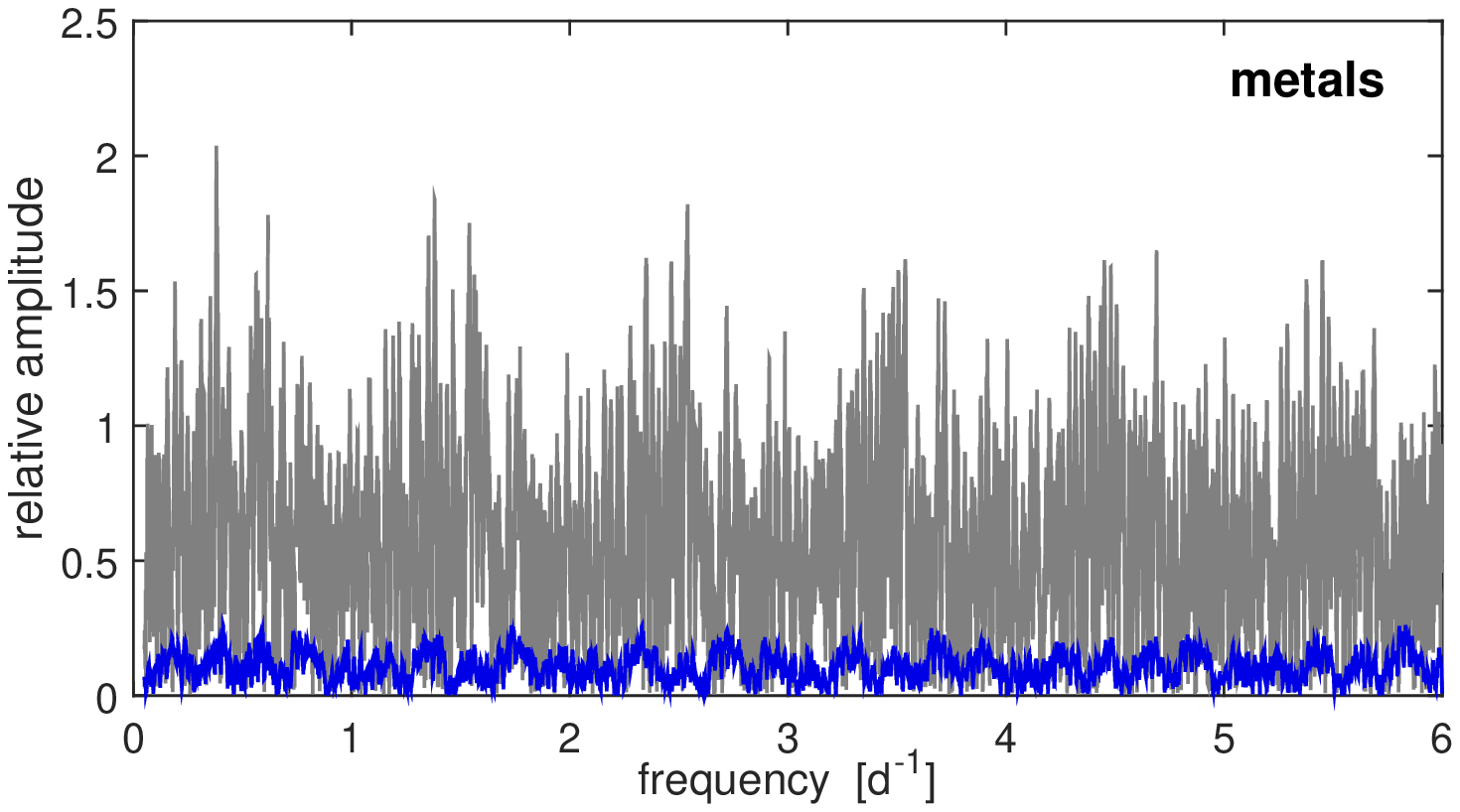}
\includegraphics[width=0.55\textwidth]{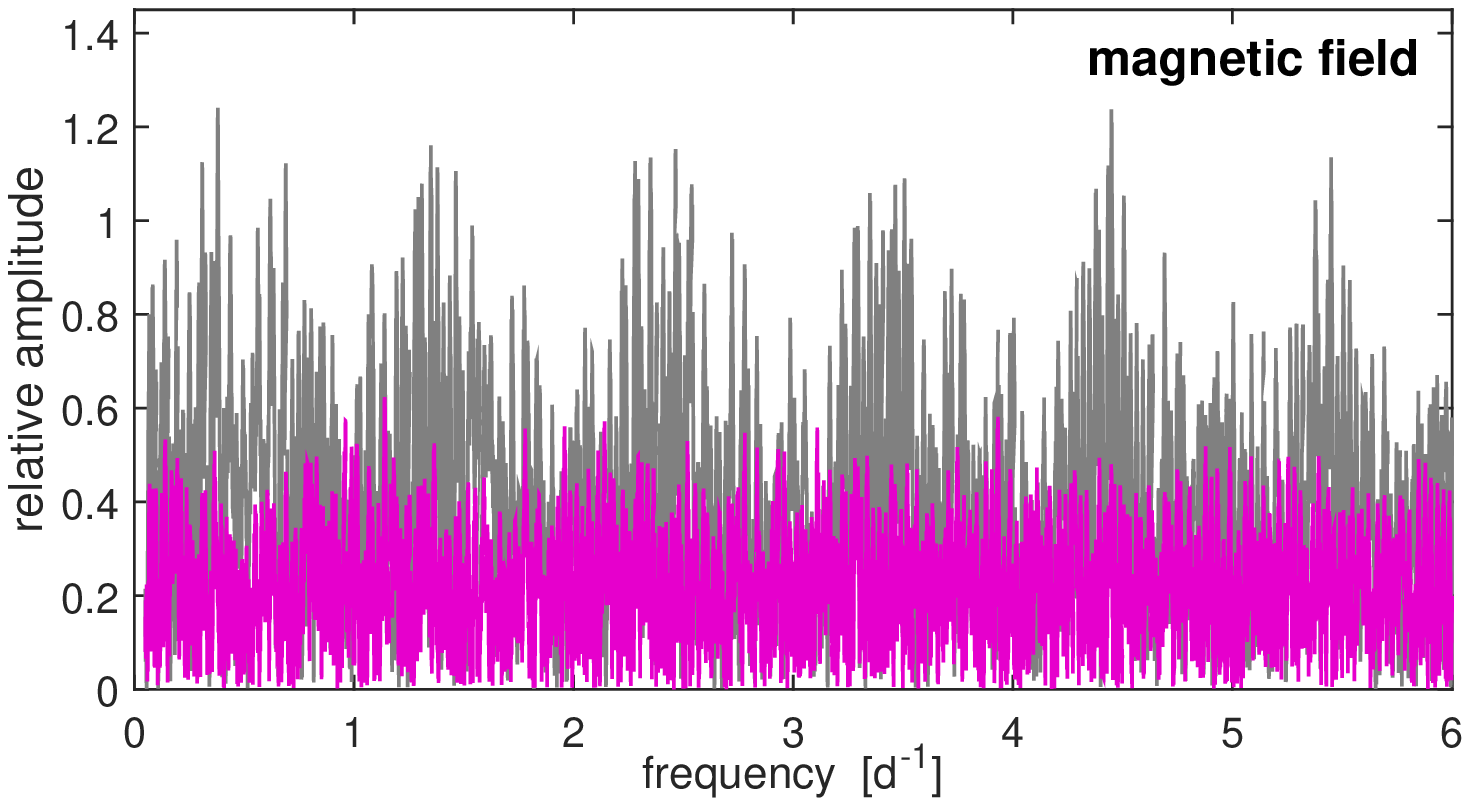}
\caption{
Periodograms (in d$^{-1}$) for the EWs of
the \ion{H}{i} lines,
the \ion{He}{i} lines,
the metallic (\ion{C}{ii}, \ion{N}{ii}, \ion{O}{ii}, and \ion{Si}{iii}) lines,
and the $\bz{}$ measurements (from top to bottom).
The differences between the observations and the model of the rotationally modulated variations
are plotted in colour in the online version of the article.}
\label{fig:hyhemag}
\end{figure}

The periodograms spanning the frequency interval from 0 to 6\,d$^{-1}$ displayed in Fig.~\ref{fig:hyhemag} prove
that the observed frequency spectra of the EWs of the \ion{H}{i}, \ion{He}{i} and metal lines, and the $\bz{}$ measurements
can be explained by rotational modulation with one period.
The whole forest of peaks are aliases of the basic frequency at $f_1=0.3805$\,d$^{-1}$.
The differences between the observations and the model of rotationally modulated variations
plotted in Fig.~\ref{fig:hyhemag} do not indicate any other periodicity,
e.g.\ the periodicity of the order of hours revealed in the photometry and discussed in Sect.~\ref{sect:fotovar}.

Using the linear ephemeris of the rotational modulation derived from the photometric observations
with the parameters $M_{0V_{\rm{max}}}=2\,453\,428.40(8)$ and $P_V=2\fd628\,07(24)$,
we can plot the phase curves of the EWs of the lines studied as well as $\bz{}$.
We found that all phase curves can be well approximated by simple sinusoids with extrema near the phases 0 and 0.5.
It appears that the maximum of the $V$ light curve coincides with the maxima of the intensity of the H, C, N, O, and Si lines
and the minima of the intensities of the He lines and the $\bz{}$ curves.

We can build a simple phenomenological model of the observed variability derived from the FORS\,2 spectra:

\begin{eqnarray}
& \displaystyle EW_i(t)=EW_{0\,i}+\frac{B_{i}}{2}\cos(2\pi\,\vartheta_{\rm S});\quad i=1,\ldots,9,\\
& \displaystyle \langle B_z\rangle_j(t)= \langle B_z\rangle_{0j}
+\frac{C}{2}\cos(2\pi\,\vartheta_{\rm S}),  j={\rm all},{\rm hyd};\ \vartheta_{\rm S}=\frac{t-M_{0\mathrm S}}{P_{\rm S}},\nonumber
\end{eqnarray}

\noindent
where $EW_i(t)$ is a model prediction of the EW of the $i-$th spectral line at the time $t$,
$EW_{0i}$ is its mean EW,
and $B_{i}$ is the amplitude of its variation.
$\bz{}_j(t)$ is a model prediction of the $\bz{}$ value for the case of measurements of all lines $(j={\rm all})$,
or only the hydrogen lines $(j={\rm hyd})$, $C$ is the amplitude of the $\bz{}$ variations,
assuming that the amplitudes of these changes for both type of measurements are the same.
$\vartheta_{\rm S}$ is the corresponding phase function,
$P_{\rm{S}}$ the period of the spectral variations,
$M_{0\rm S}$ is the time of helium and $\bz{}$ minimum and the intensity maximum of the other chemical elements.
The spectral model solved by the weighted least square method then leads to the following ephemeris parameters:
$P_{\rm S}=2\fd627\,60(25)$ and $M_{0\rm S}=2\,457\,465.173(6)$.

\begin{table}
\begin{center}
\caption{
Ephemerides of \CPD{} resulting from the different data.
}
\label{tab:fotper}
\centering
\begin{tabular}{lccc}
\hline
\hline
\multicolumn{1}{c}{} &
\multicolumn{1}{c}{Photometry} &
\multicolumn{1}{c}{Spectroscopy and} &
\multicolumn{1}{c}{All data} \\
\multicolumn{1}{c}{} &
\multicolumn{1}{c}{} &
\multicolumn{1}{c}{Spectropolarimetry} &
\multicolumn{1}{c}{} \\
\hline
  $P_1$    & 2\fd628\,07(24)     & 2\fd627\,60(25)    & 2\fd628\,09(5)     \\
  $M_{01}$ & 2\,453\,428.40(8)   & 2\,457\,465.173(6) & 2\,457\,444.146(8) \\
  $P_2$    & 0\fd409\,879(6)     & & \\
  $M_{02}$ & 2\,453\,429.488(12) & & \\
  $P_3$    & 0\fd333\,617(5)     & & \\
  $M_{03}$&2\,453\,429.777(10)   & & \\
\hline
\end{tabular}
\end{center}
 \end{table}

The spectroscopically found period $P_{\rm S}$ is fully compatible with the one derived from photometry, $P_V$.
The prediction of the time of $M_{0\rm S}$ from photometry is 2\,457\,465.12(4),
which is in a satisfactory agreement with our finding that the flux maxima coincide with the maxima of H, C, N, O, and Si.
Assuming that $P_V=P_{\rm S}=P_1$ and $M_{0V}=M_{0\rm S}=M_{01}$,
we can establish a general model of the photometric, spectroscopic and spectropolarimetric variations of \CPD{} with a solid time basis.
Applying this model to the full set of observational data, we found the following ephemeris:
$P_1=2\fd628\,09(5)$, $M_{01}=2\,457\,444.146(8)$.
The periods and ephemerides found from the photometric data are
$P_2=0\fd409\,879(6)$, $M_{02}=2\,453\,429.488(12)$,
$P_3=0\fd333\,617(5)$, and $M_{03}=2\,453\,429.777(10)$
(see also Table~\ref{tab:fotper}).

\section{Rotationally modulated variations}

\subsection{Inclination of the rotation axis}

Knowing the rotation period $P_1$ we can also estimate the equatorial velocity $v_{\rm {eq}}$
and the inclination of the rotation axis of the star $i$.
Assuming the radius $R=5.4\pm1.0$\,R$_{\odot}$ and the projection
of the equatorial rotation velocity $v\sin i=35\pm5$\,km\,s$^{-1}$,
we obtain for the equatorial velocity $v_{\rm {eq}}=50.6\ R/P_1=100\pm20$\,km\,s$^{-1}$,
and the rotation axis inclination $i=20^{\circ}\pm5^{\circ}$.
\CPD{} is thus a nearly pole-on star.

\subsection{Magnetic field geometry}

\begin{figure}
\centering
\includegraphics[width=0.52\textwidth]{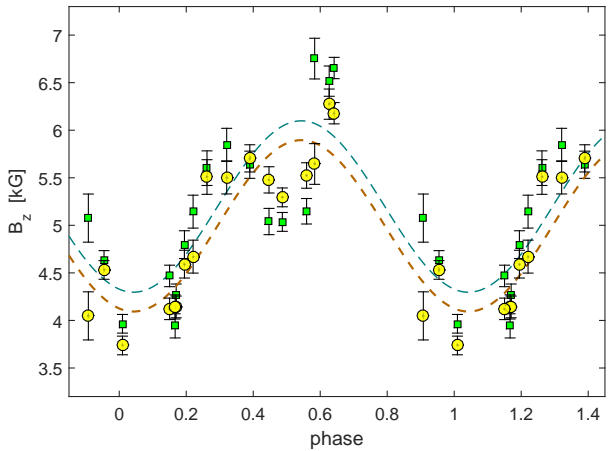}
\caption{
Phase curves for the $\bz{}_{\rm{all}}$ and $\bz{}_{\rm{hyd}}$ magnetic measurements
(circles and squares, respectively).
The sinusoidal fits through the measurements are denoted by the lower and upper dashed lines, respectively.
}
\label{fig:bzfaze}
\end{figure}

The detailed bootstrapping analysis of the $\bz{}$ measurements shows that the phase curves
derived from the spectropolarimetric measurements of only the hydrogen or all spectral lines can be approximated
by the same sinusoid with an amplitude of $1.85\pm0.24$\,kG with mean values of $\bz{}_{0,{\rm hyd}}=5.14\pm0.15$\,kG
and $\bz{}_{0,{\rm all}}=4.93\pm0.09$\,kG, respectively (see Fig.~\ref{fig:bzfaze}).
The minimum of the mean longitudinal magnetic field takes place at the phase $(0.045\pm0.025)$.
The extreme values of $\bz{}_{\rm{all}}$ are $\bz{}^{\rm{max}}_{\rm{all}}=5.85\pm0.15$\,kG
and $\bz{}^{\rm{min}}_{\rm{all}}=4.01\pm0.15$\,kG.

Using the well known relations developed by \citet{Stibbs1950} and \citet{Preston1967}
for a centred magnetic dipole tilted to the rotation axis by angle $\beta$,
we find $\beta=28^{\circ}\pm7^{\circ}$.
The uncertainty of this angle is mainly due to the uncertainty in the knowledge of the true radius of the star.
Nevertheless,
because of $\cos(\beta-i)\approx 1$,
we conclude that the sum $i+\beta=47^{\circ}\pm3^{\circ}$ is almost independent of the choice of the radius of the star.
Assuming a linear limb-darkening coefficient $u=0.3$,
typical for the effective temperature $T_{\rm{eff}}=23\,650$\,K \citep{Castro2017},
we can estimate that the dipole strength $B_{\rm p}$ is 21\,kG.

We observe that the phase curve of $\bz{}$ displays a secondary minimum at phase 0.5,
indicating the possibility of a more complex magnetic field with a non-zero
quadrupole component (see Fig.~\ref{fig:bzfaze}).
We note that complex global magnetic fields are frequent in hot He-rich stars with low age.
In any case, the dipole component seems to be dominant in \CPD{}, so that the analysis presented above remains valid.

\subsection{Spectral variations}

17 low-dispersion FORS\,2 polarimetric spectra do not allow to carry out detailed Zeeman Doppler imaging of the distribution
of chemical elements on the stellar surface.
On the other hand, they can contribute to our understanding of the stellar rotationally modulated variability.
Variations in EWs are caused by an inhomogeneous distribution
of certain chemical elements on the surface of a rotating star and
the behaviour of the EW phase curves is determined by the location of spots
with enhanced element abundances and the inclination of the rotation axis $i$.
For small inclination angles $i<30^{\circ}$,
we observe only structures near the visible rotation pole and a considerable part of the chemical spots are
hardly visible or even invisible.
This is why the EW phase curves of nearly pole-on stars show single-waves and are almost symmetrical.

\begin{figure}
\centering
\includegraphics[width=0.45\textwidth]{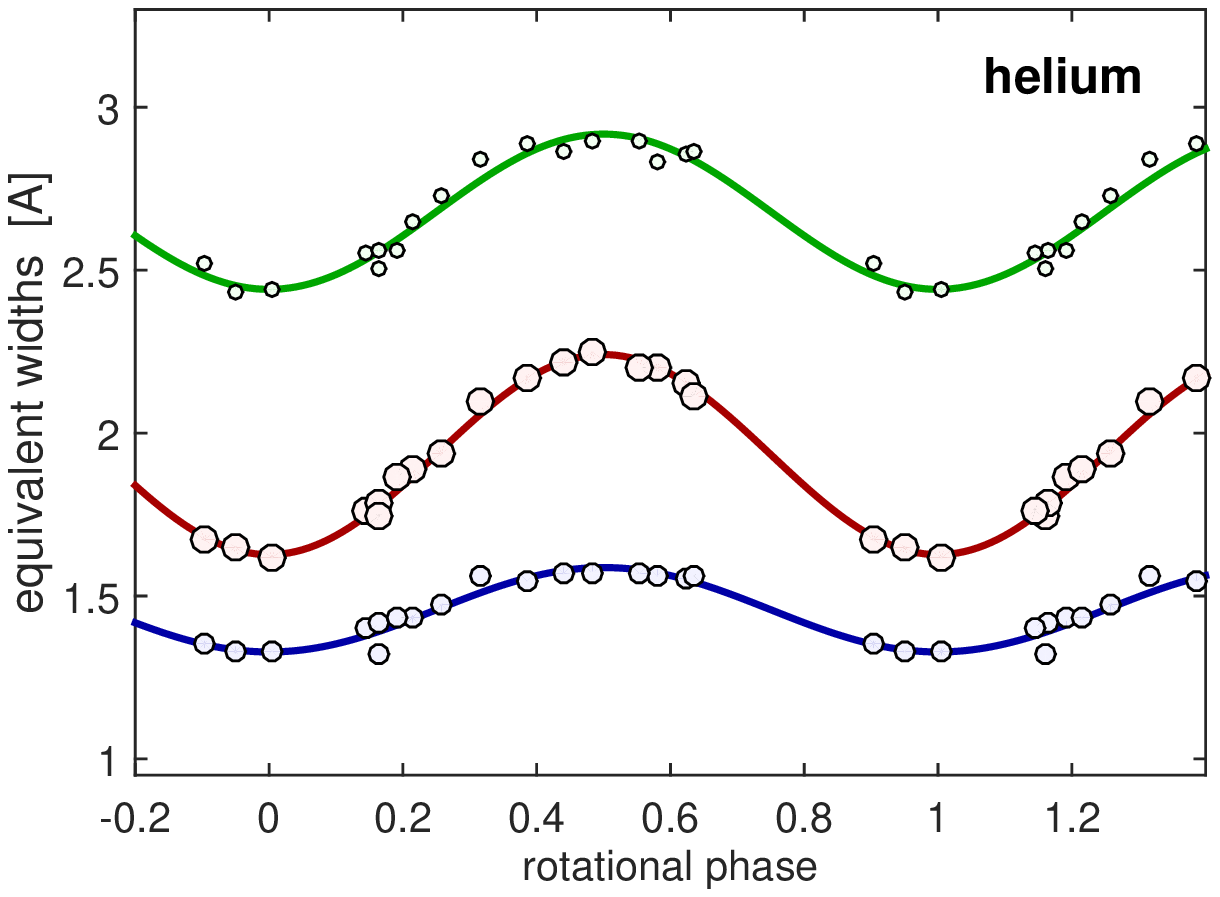}
\includegraphics[width=0.45\textwidth]{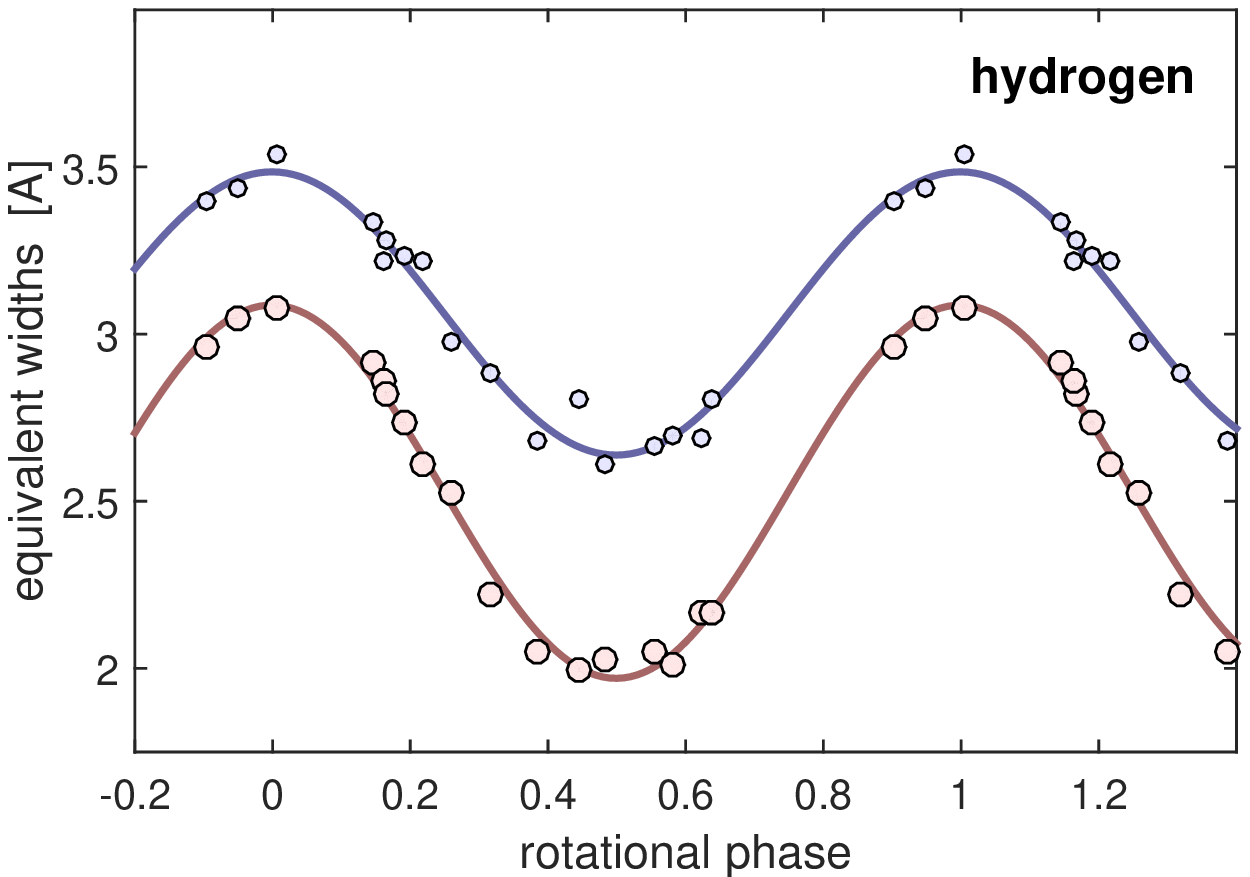}
\caption{
{\sl Upper panel:}
EWs of three \ion{He}{i} lines plotted versus the rotation phase.
The sinusoidal fits with minima near phase zero are denoted by lines.
Upper line - \ion{He}{i}~4471,
middle line - \ion{He}{i}~4388,
and lower line - \ion{He}{i}~4922.
{\sl Lower panel:}
EWs of two Balmer lines plotted versus the rotation phase.
The sinusoidal fits with maxima near phase zero are denoted by lines.
Upper line - H$\gamma$ and lower line - H$\beta$.
The areas of the circles are proportional to their weights.
}
\label{fig:hehfaze}
\end{figure}

As a rule, the distribution of helium on the surface of He-rich stars is very uneven.
The abundance of helium in spots has to be much larger than outside of spots, causing
the observed strong variations in the EWs of the helium lines.
The phase curves of all three measured lines (\ion{He}{i}~4471, 4388, and 4492) display strong,
well-defined symmetric variations
(see the upper panel in Fig.~\ref{fig:hehfaze}). This is compatible with our concept of the presence of one helium spot
on the surface of CPD\,$-$62$^{\circ}$2124.
The minimum of the EW of the \ion{He}{i} spectral lines takes place at the phase $\varphi=-0.001\pm0.004$.

The lower panel in Fig.~\ref{fig:hehfaze} demonstrates that the EW of the hydrogen Balmer
H$\beta$ and H$\gamma$ lines vary in antiphase with the \ion{He}{i} lines.
The maximum of the EW of the \ion{H}{i} spectral lines takes place at $\varphi=0.002\pm0.004$.

\begin{figure*}
\centering
\includegraphics[width=0.245\textwidth]{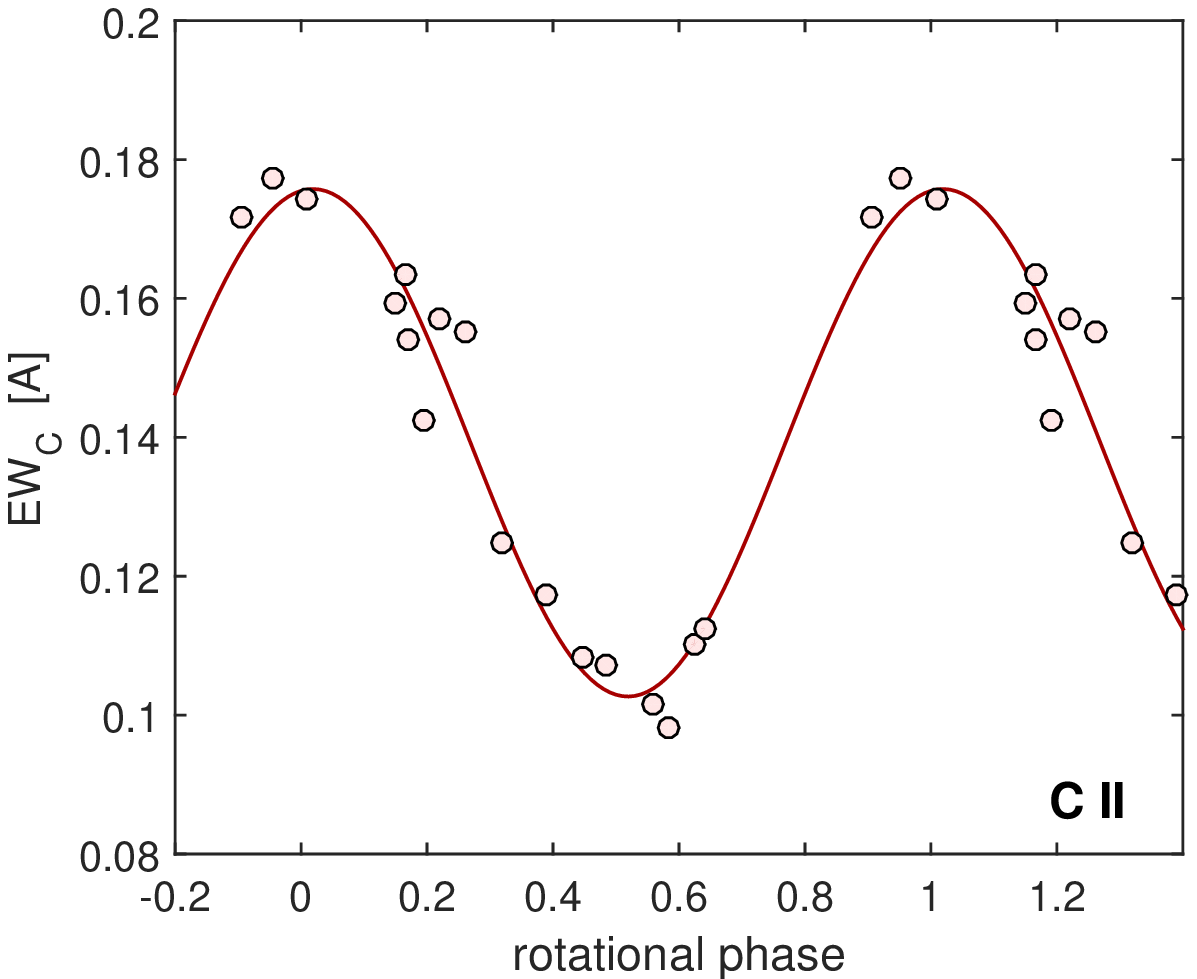}
\includegraphics[width=0.245\textwidth]{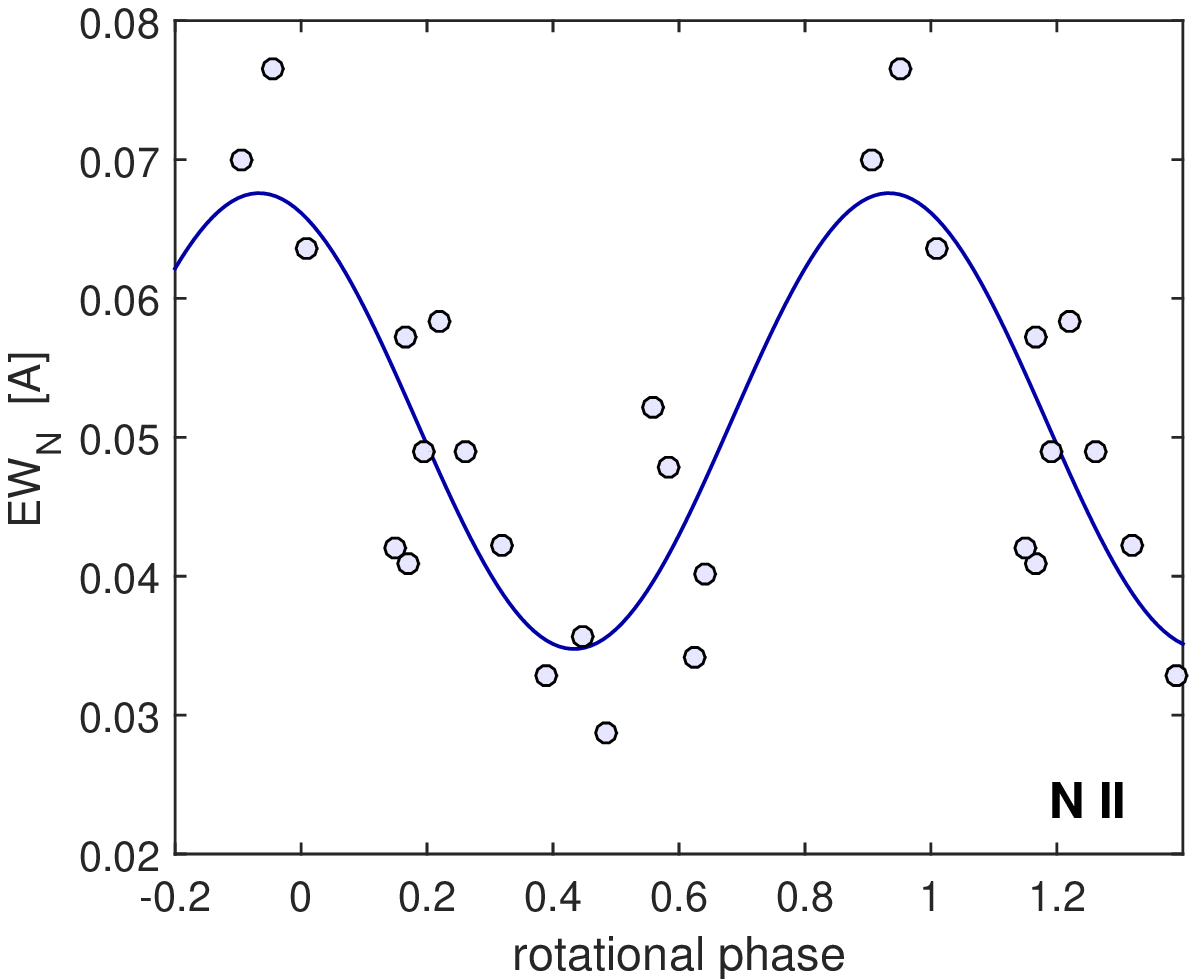}
\includegraphics[width=0.245\textwidth]{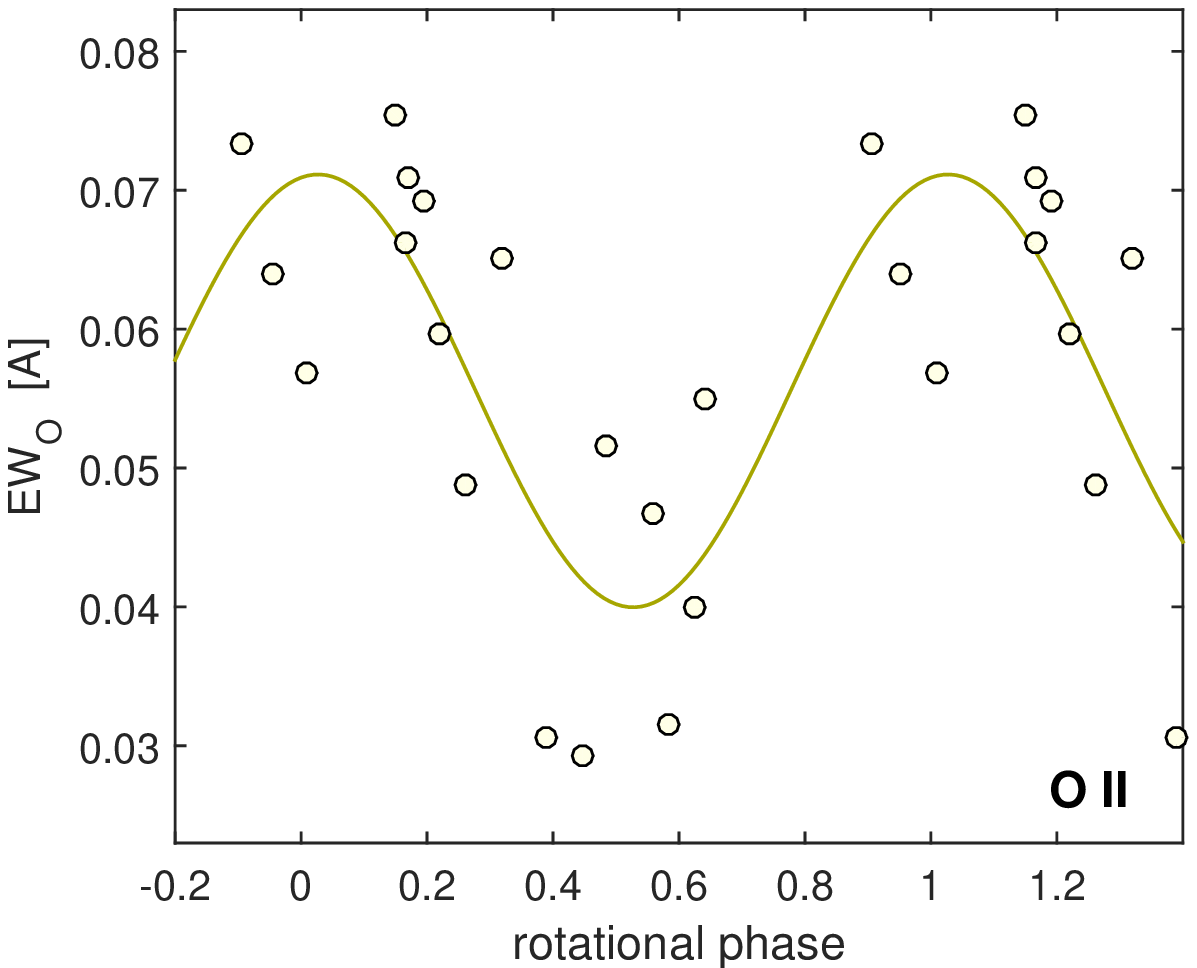}
\includegraphics[width=0.245\textwidth]{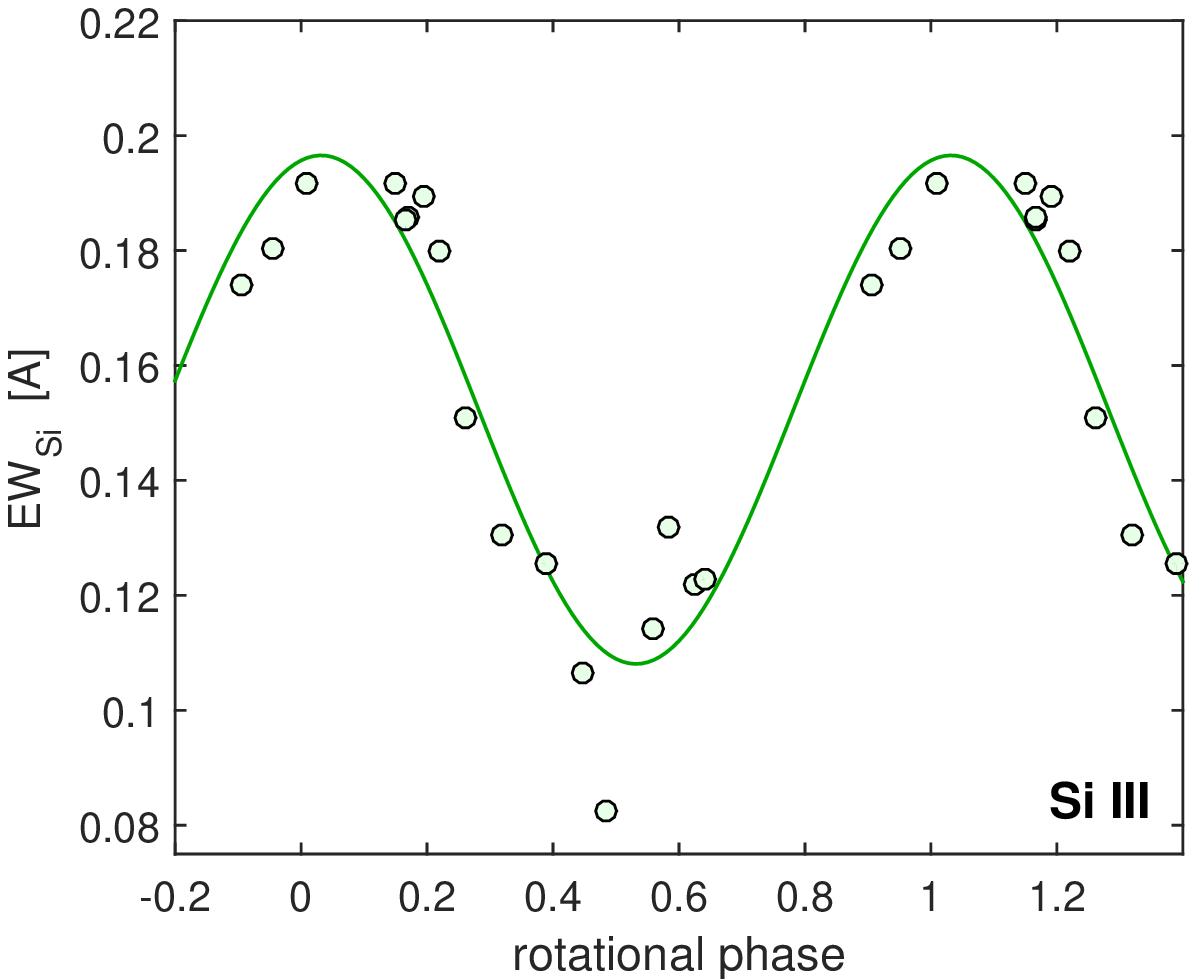}
\caption{
Equivalent widths (in \AA{}) of
the \ion{C}{ii}~4267, \ion{N}{ii}~4631, \ion{O}{ii}~4662, and \ion{Si}{iii}~4553 lines
plotted versus the rotation phase.
The solid lines indicate the sinusoidal fits, where
all phase curves have their maxima near the zero phase.
The largest relative variations in EW exhibit lines of silicon,
suspected to be a source of rotationally modulated flux variations.
}
\label{fig:metalfaze}
\end{figure*}

The same can be said about the EWs of the metal lines belonging to \ion{C}{ii}, \ion{N}{ii}, \ion{O}{ii},
and \ion{Si}{iii}.
All phase curves can be well fitted by simple sinusoids with a maximum at phase 0 (see Fig.~\ref{fig:metalfaze}).

\subsection{Rotationally modulated flux variation and photometric pulsations}

Our frequency analysis of 844 detrended ASAS3 measurements revealed three independent periodicities
characterized by their periods $P_1,P_2,\ \mathrm{and}\ P_3$.
The last two periods of the order of hours can probably be referred to the presence of pulsations, while the first
period is very likely the rotation period.
Since all amplitudes of the flux variations can be attributed to periodic mechanisms,
we consider now the rotational periodicity by removing the other two periodicities.

\begin{figure}
\centering
\includegraphics[width=0.45\textwidth]{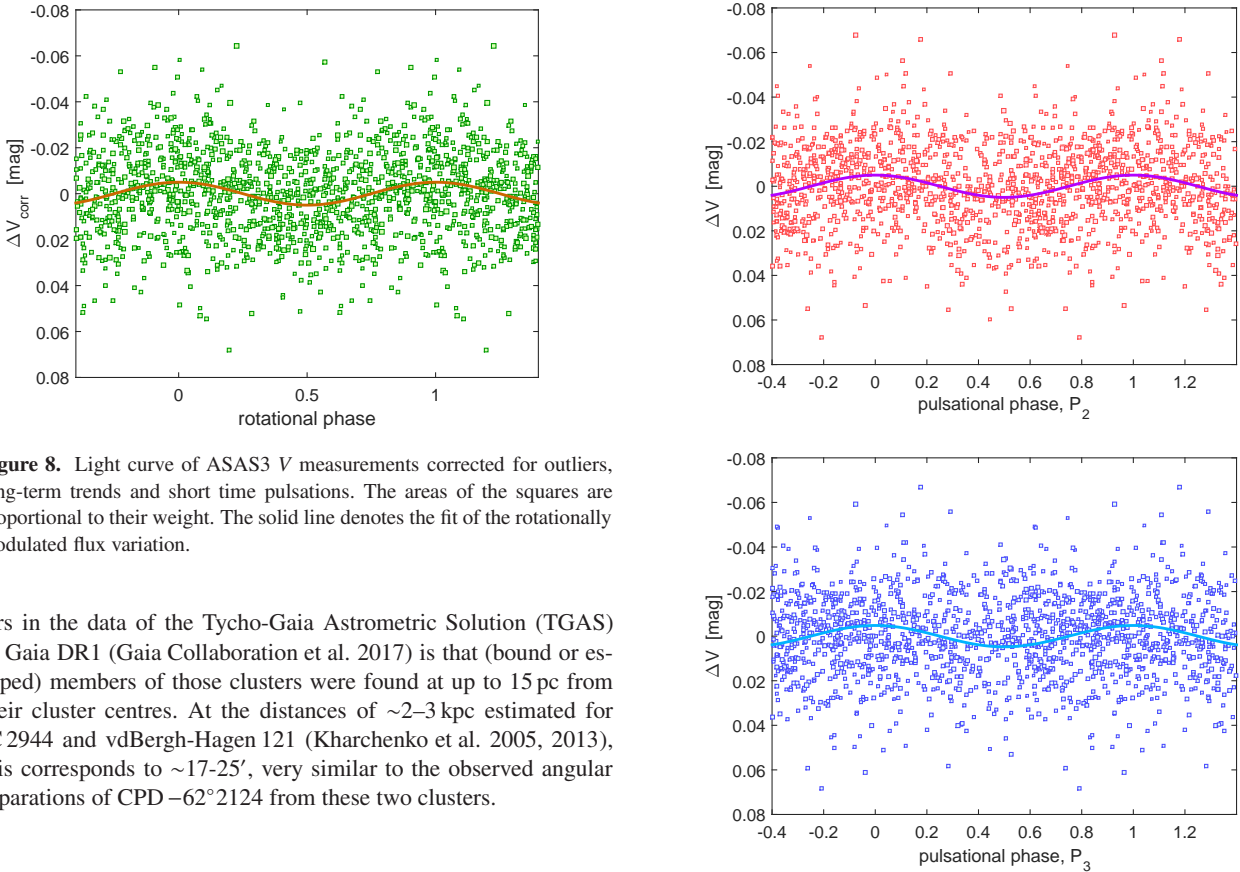}
\caption{
Light curve of ASAS3 $V$ measurements corrected for outliers, long-term trends and short time pulsations. The areas of the
squares are proportional to their weight. The solid line denotes the fit of the rotationally modulated flux variation.
}
\label{fig:vp1faze}
\end{figure}

In Fig.~\ref{fig:vp1faze}, the maximum of the light curve appears at phase 0 and the amplitude of the
flux phase curve is $0.0099\pm0.0018$\,mag.
The variations are caused by the presence of a photometric spot, bright in $V$,
whose photocenter passes the meridian at the phase of $-0.009\pm0.029$.
Its brightness is very probably the result of a redistribution of radiative energy
from the ultraviolet into the visible due to the back-warming process in regions with overabundant elements possessing
numerous lines in the ultraviolet.
Silicon is the element mainly
responsible for flux variability in HD\,37776 \citep{krt07} and other
hot CP stars \citep{krt15}.

The relatively small amplitude of the flux variability is due to two causes:
a) The contrast of the photometric spot located outside of the helium spot is lowered
because also in helium spots the 
redistribution of energy from ultraviolet to optical regions of the spectrum takes place,
but not so efficiently as in e.g.\ silicon spots,
b) the photometric spot tracks during the rotational cycle within the central part of the visible hemisphere of the star,
so that its projection change is small.

\begin{figure}
\centering
\includegraphics[width=0.45\textwidth]{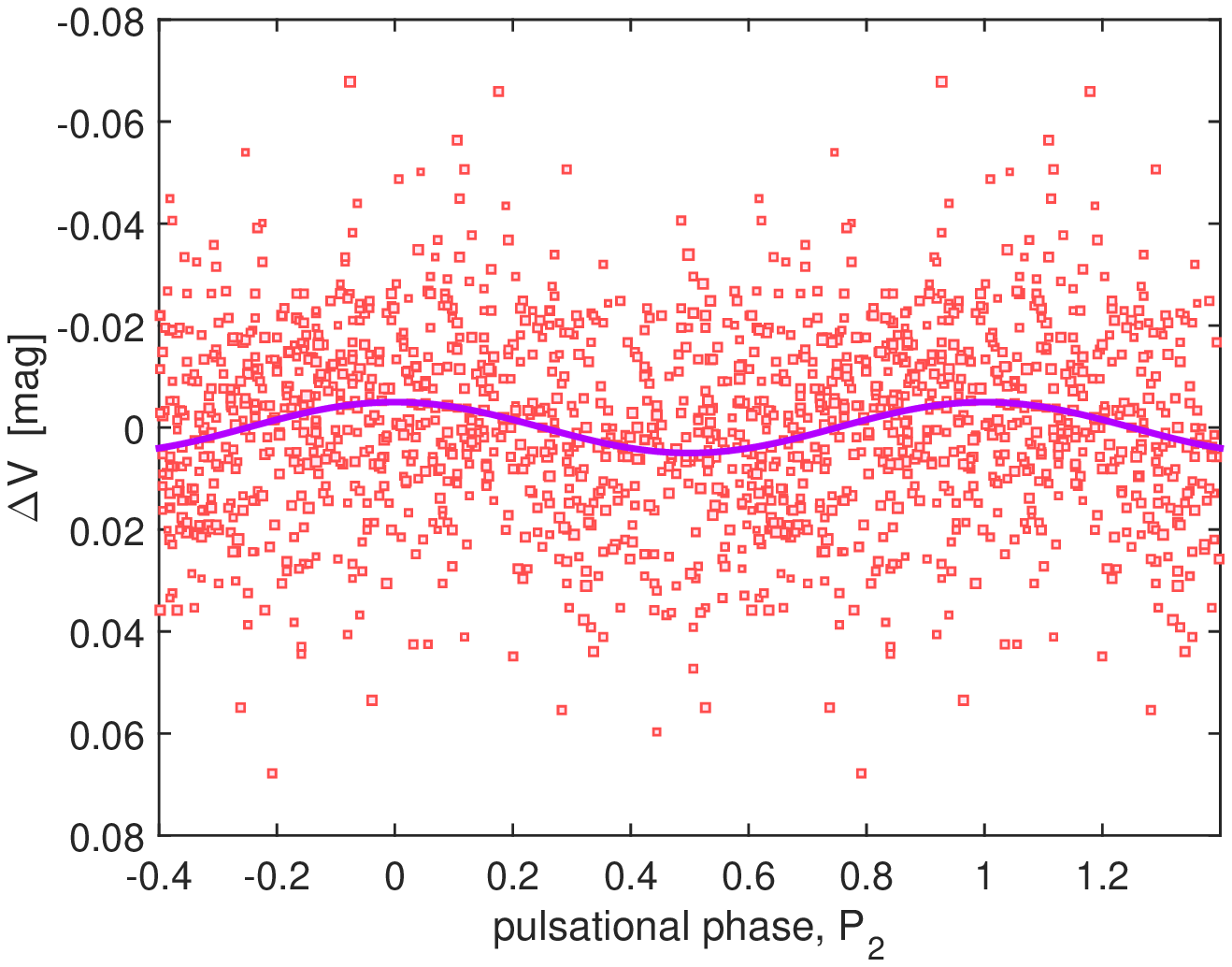}
\includegraphics[width=0.45\textwidth]{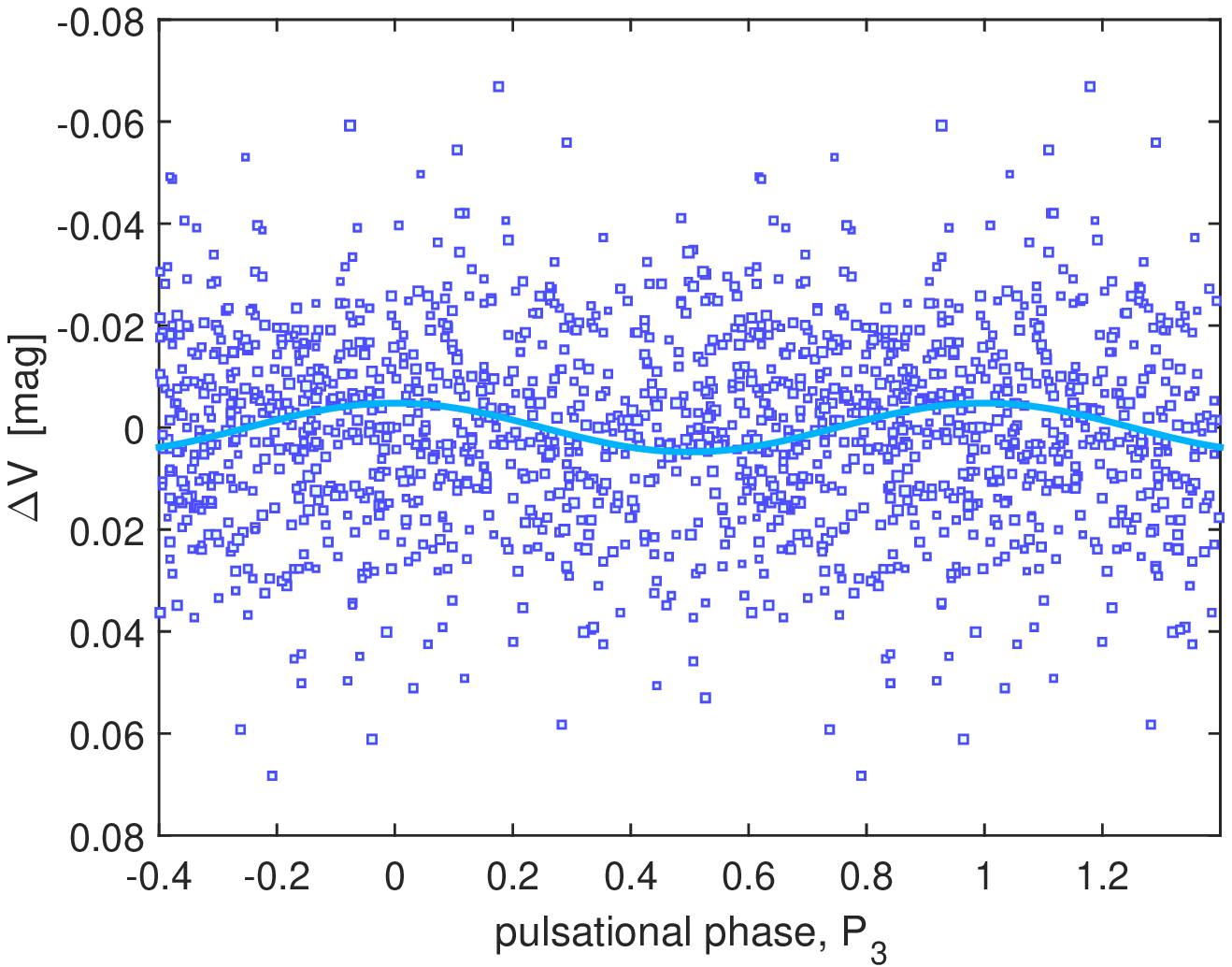}
\caption{
{\sl Upper panel:}
$V$ measurements versus the pulsational phase for the period $P_2=0\fd409\,879$.
{\sl Lower panel:}
$V$ measurements versus the pulsational phase for the period $P_3=0\fd333\,617$.
The area of the individual markers is proportional to their weight.
}
\label{fig:pulsfaze}
\end{figure}

The photometry of \CPD{} reveals two other frequencies higher than 1\,d$^{-1}$,
which we can probably attribute to its pulsations.
The light curves of the proposed pulsations are sinusoids with relatively small amplitudes of 0.01\,mag
(see Fig.~\ref{fig:pulsfaze}).
The pulsations are apparent only in the photometric frequency spectrum. Based on the FORS\,2 observations
it seems that they can have an impact on the line profile shapes. However, they do neither significantly influence the
intensities of the spectral lines, nor the global magnetic field.

\section{Evolutionary state}

To better understand the evolutionary status of CPD\,$-$62$^{\circ}$2124,
we also exploited currently available astrometric data.
\citet{dias14} investigated stars in open cluster areas
using proper motions from the fourth US Naval
Observatory CCD Astrograph Catalog (UCAC4;
\citealt{zacharias13}) and found a 98\% membership
probability for CPD\,$-$62$^{\circ}$2124 being a member of the open cluster IC\,2944.
However, they gave an angular distance to the cluster centre of 19.4\arcmin{}.
This is larger than the cluster radius of 14.4\arcmin{}, determined in the
cluster survey of \citet{kharchenko05}, based on
Hipparcos and Tycho-2 \citep{esa97,hog00}
proper motions and optical photometry, and of 12\arcmin{} in the more recent
survey of \citet{kharchenko13}, based on proper motions
from the PPMXL catalogue \citep{roeser10} and
near-infrared photometry from the Two Micron All Sky Survey (2MASS;
\citealt{skrutskie06}). The most accurate new proper motion
measurements of CPD\,$-$62$^{\circ}$2124 involving Gaia DR1 data
\citep{gaia16a,gaia16b,lindegren2016}
in the Hot Stuff for One Year (HSOY; \citealt{altmann17})
and UCAC5 \citep{zacharias17} catalogues yield
($\mu_{\alpha}\cos{\delta}$, $\mu_{\delta}$) of
($-$7.7$\pm$1.1, $+$2.0$\pm$1.1)\,mas/yr and
($-$7.3$\pm$1.0, $+$0.2$\pm$1.0)\,mas/yr, respectively.

There seems to be a good agreement (especially of the UCAC5 values)
with the mean IC\,2944 proper motion of
($-$6.2$\pm$0.9, $-$1.2$\pm$0.9)\,mas/yr
from \citet{kharchenko05} and
($-$7.0$\pm$0.9, $-$1.1$\pm$0.9)\,mas/yr
from \citet{kharchenko13}.
However, there is a second cluster, vdBergh-Hagen\,121, with an only
slightly different proper motion of
($-$5.1$\pm$0.5, $+$1.1$\pm$0.6)\,mas/yr and
($-$5.9$\pm$0.8, $+$0.5$\pm$0.8)\,mas/yr
listed in the latter two catalogues, respectively.
The angular distance of CPD\,$-$62$^{\circ}$2124 to the cluster
vdBergh-Hagen\,121 is again larger than the cluster radius, but a bit
smaller ($\sim$17.2\arcmin{}) than to the cluster IC\,2944.
Because of the large angular distances from both cluster centres,
\citet{kharchenko13} considered CPD\,$-$62$^{\circ}$2124
as a spatial non-member in these two clusters, but its membership
probabilities from proper motion, $JK_s$, and $JH$ photometry are
higher with respect to IC\,2944 (36\%, 100\%, 100\%) than
to vdBergh-Hagen\,121 (17\%, 37\%, 100\%).

The question, whether CPD\,$-$62$^{\circ}$2124 belongs to a star
cluster (and to which one), can probably be answered only when Gaia DR2
data will become available in 2018 April. So far, we can only speculate on
this issue, mainly because of the large angular distance of the star
to its potential birth clusters. As open clusters dissolve after their
formation, one can further speculate on the possible status of
CPD\,$-$62$^{\circ}$2124 as a former cluster member, which represents
the oldest generation of stars that formed in the cluster.
An interesting new result from an investigation of nearby ($\lessapprox$450\,pc)
open clusters in the data of the Tycho-Gaia
Astrometric Solution (TGAS) of Gaia DR1 \citep{gaia17}
is that (bound or escaped) members of those clusters were found at up to
15\,pc from their cluster centres.
At the distances of $\sim$2--3\,kpc
estimated for IC\,2944 and vdBergh-Hagen\,121
\citep{kharchenko05,kharchenko13}, this corresponds to
$\sim$17-25\arcmin{}, very similar to the observed angular separations of
CPD\,$-$62$^{\circ}$2124 from these two clusters.

\section{Discussion}
\label{sect:disc}

We exploited low-resolution spectropolarimetric observations taken by FORS\,2 and 844 ASAS3 photometric measurements
for the determination of the rotation period, pulsations, and the magnetic field geometry of \CPD{}. 
For the analysis of the available data, 
we applied periodograms and phenomenological models of flux, spectral and spectropolarimetric variability.
Line intensities belonging 
to several elements, the mean longitudinal magnetic field $\bz{}$, and flux in the $V$ filter, were found to vary with 
the same period $P_1 = 2.628\,09(5)$\,d, which was identified as the rotation period.
The magnetic field of CPD\,$-$62$^{\circ}$2124 is
dominated by a dipolar component of 21\,kG, tilted to the rotation axis by the angle $\beta=28^{\circ}\pm7^{\circ}$, while the 
inclination of the rotation axis is only $i=20^{\circ}\pm5^{\circ}$.  Such an extraordinarily strong magnetic field has never been 
detected in any massive early-B type star.
The analysis of the ASAS3 photometric measurements reveals, apart from the rotational flux modulation, also the presence of 
two additional 
periodicities of 0.409\,879(6)\,d and 0.333\,617(5)\,d, which are most probably caused by pulsational
variability.
We note that these two periods appear rather unusually long for a $\beta$~Cephei type pulsation.
It would be worthwhile to obtain extensive spectroscopic time series
to investigate the pulsational characteristics of this star in more detail.
The classification of CPD\,$-$62$^{\circ}$2124 as a pulsating He-rich star with a measured huge magnetic field 
would place it also as a record holder among all magnetic upper-main sequence pulsating $\beta$~Cep and slowly pulsating B (SPB) stars, 
which usually exhibit much lower magnetic field 
strengths \citep[e.g.,][]{Hubrig2006,Hubrig2009}.

Remarkably, this strongly magnetic star, although being in an advanced evolutionary state and having finished 
approximately 60\% of its main-sequence life, rotates rather fast with $v_{\rm eq}$ of about 100\,km\,s$^{-1}$ and has a
rather short rotation period of 2.628\,d. It is of interest that while for the majority of the magnetic Ap and Bp stars it is 
expected that magnetic braking slows stellar rotation \citep[e.g.,][]{Mathys2004}, the study of a representative sample of 
this type of stars with strong magnetic fields 
by \citet{Hubrig2000} did not show any correlations between the
rotation period and the fraction of the main-sequence lifetime completed.
Clearly, strongly magnetic early-B type stars in an advanced evolutionary state are of special interest for our 
understanding of the evolution of the angular momentum and of spindown timescales in the presence of a magnetic field. 

Our study of the spectral variability shows the presence of significant chemical
abundance variations across the stellar photosphere with the helium abundance decreasing towards the
longitudinal magnetic field minimum and an increase of the abundances of other elements at the same phase.
Future high-resolution high signal-to-noise spectropolarimetric observations will be worthwhile to determine
the surface distribution of the different elements.
Since the discovered pulsational variability is most likely caused by $\beta$~Cep-like pulsations,
a detailed seismic study of CPD\,$-$62$^{\circ}$2124 will be of
great importance to constrain the physical processes in magnetic stars and to test the impact of the magnetic field on the
internal mixing processes.
As this star is rather bright and has a relatively low projected rotation velocity, it
appears to be an excellent candidate also for future high-resolution polarimetric analyses to investigate various
atmospheric effects that interact with a strong magnetic field.

\section*{Acknowledgments}
\label{sect:ackn}

We thank the anonymous referee for the constructive comments. Further, we acknowledge a fruitful discussion 
with J.~Krti\v{c}ka.
Based on observations made with ESO Telescopes at the La Silla Paranal Observatory under 
programmes 097.D-0428(A) and 191.D-0255(G). AK acknowledges financial support from the RFBR grant 16-02-00604A. 
ZM and MZ acknowledge support from the GA\v{C}R\,16-01116S project.

\label{lastpage}

\end{document}